\documentclass[12pt]{article}

\input epsf
\usepackage{amssymb}
\usepackage[dvips]{graphicx}
\usepackage{amscd}
\usepackage{mathrsfs}
\usepackage{amsmath}

\newcommand{\be}{\begin{equation}}
\newcommand{\ee}{\end{equation}}
\newcommand{\beqa}{\begin{eqnarray}}
\newcommand{\eeqa}{\end{eqnarray}}
\newcommand{\nn}{\nonumber}

\begin{document}
\linespread{1.5} \flushbottom

\begin{center}
{\Large\bf{ Coulomb problem in NC quantum mechanics:\\
Exact solution and non-perturbative aspects }} \vskip1cm
V. G\' alikov\' a and P. Pre\v{s}najder\\
{\it Faculty of Mathematics, Physics and Informatics,\\
Comenius University Bratislava, Slovakia}
\end{center}

\begin{abstract}
The aim of this paper is to find out how would possible space non-commutativity (NC) alter the QM solution of the Coulomb problem. The NC parameter $\lambda$ is to be regarded as a measure of the non-commutativity - setting $\lambda = 0 $ means a return to the standard quantum mechanics.
As the very first step a rotationaly invariant NC space ${\bf R}^3_\lambda $,  an analog of the Coulomb problem configuration space ${\bf R}^3_0\,=\,
{\bf R}^3\setminus \{0\}$, is introduced.
${\bf R}^3_\lambda $ is generated by NC coordinates realized as operators acting in an auxiliary (Fock) space ${\cal F}$.
The properly weighted Hilbert-Schmidt operators in ${\cal F}$ form  ${\cal H}_\lambda  $, an NC analog of the Hilbert space of the wave functions. We will refer to them as "wave functions" \, also in the NC case. The definition of an NC analog of the hamiltonian as a hermitian operator in  ${\cal H}_\lambda  $ is one of the key parts of this paper.
The resulting problem is exactly solvable. The full solution is provided, including formulas for the bound states for $E < 0$ and low-energy scattering for $E > 0$ (both containig NC corrections analytic in $\lambda$) and also formulas for high-energy scattering and unexpected bound states at ultra-high energy (both containing NC corrections singular in $\lambda$). All the NC contributions to the known QM solutions either vanish or disappear in the limit $\lambda\to 0$.

\end{abstract}

\newpage

\section{Introduction}

Basic ideas of non-commutative geometry have been developed in \cite{Con}
and, in a form of matrix geometry, in \cite{Mad1}. The main applications
have been considered

$\bullet $ in the area of quantum quantum field theory in order to understand,
or even to remove, UV singularities, and  eventually,

$\bullet $ to formulate a proper base for the quantum gravity.

The analysis performed in \cite{DFR} led to the conclusion that quantum
vacuum fluctuations and Einstein gravity could create  (micro)black holes
which prevent localization of space-time points. Mathematically this
requires non-commutative (NC) coordinates $x^\mu$ in space-time satisfying
specific uncertainty relations. The simplest set of NC coordinates $x^\mu$
should satisfy Heisen\-berg-Moyal commutation relations
in an auxiliary Hilbert space
\be\label{DFR1} [x^\mu,x^\nu]\ =\ i\,\theta^{\mu\nu},\ \
\mu,\nu = 0,1,2,3\,,\ee
where $\theta^{\mu\nu}$ are given numerical constants that specify the
non-commutativity of the space-time in question.

Later in \cite{Jab} it was shown that field theories in NC spaces
with (\ref{DFR1}) can emerge as effective low energy limits of
string theories. These results supported a vivid development of
non-commutative QFT. However, such models contain various
unpleasant and unwanted features. The divergences are not removed,
on the contrary, UV-IR mixing appears, \cite{UVIR}. The Lorentz
invariance is broken down to $ SO(2)\times SO(1,1)$, but even this is
sufficient to prove the classical CPT and spin-statisics theorems,
\cite{CPT}. This was not accidental and led to the twisted Poincar\'e
reinterpretation of NC space-time symmetries, \cite{TwP}.
%One of the most remarkable results in this area is the proof of existence
%of a non-trivial 4 dimensional Euclidean NC QFT for self-interacting scalar
%field with $\phi^4 +(\theta^{-1} x)^2 \phi^2$ interaction term, \cite{GR}.

However, it could be interesting to reverse the approach. Not to
use the NC geometry to improve the foundations of QFT,  what is a
very complicated task, but to test the effect of non-commutativity
of the space on the deformation of the well-defined quantum
mechanics (QM):

$\bullet $ Various QM systems have been investigated in 3D space
with Heisen\-berg-Moyal commutation relations
$[x_i,x_j] = i \theta^{ij}$, $i,j = 1,2,3$, e.g.
harmonic oscillator, Aharonov-Bohm effect, Coulomb problem, see
\cite{AB}, \cite{CP}. However, in such 3D NC space the rotational
symmetry is violated and there are systems, such as $H$-atom, that
are tightly related to the rotational symmetry.

$\bullet $ The rotational symmetry survives in 2D Heisenberg-Moyal
space with NC coordinates $x_1,x_2$ satisfying the
 commutation relations $[x_1,x_2] = i
\theta$ in an auxiliary Hilbert space. In \cite{Scholtz} a planar
spherical well was described in detail:

(i) First, the Hilbert space ${\cal H}$ of operator wave functions
$\psi  = \psi (x_1,x_2)$ was defined;

(ii) Further, the Hamiltonian was defined as an operator acting in
${\cal H}$. It was nice to see how the persisted rotational
symmetry helps to solve exactly the problem in question.

The presented list of references is incomplete and we apologize
for that. We restricted ourselves to those which initiated
progress or are close to our approach.

Our aim is to extend this scheme to the QM problems with
rotationally symmetric potentials $V(r)$ in the configuration
space $R^3_0\,\equiv\,R^3 \setminus\{0\}$. We restrict ourselves
to the Coulomb potential which, in the usual (commutative)
setting, is a solution of the Laplace equation vanishing at infinity:
\be\label{coul0} \Delta V(r)\,=\,0\ \ \Rightarrow\ \ V(r)\ =\
-\,\frac{q}{r}\,. \ee
For Coulomb problem , in a Gaussian system of units, $q$ is a square of
electric charge $q\,=\,\pm\,e^2$: $q>0$ or $q<0$ corresponding
to the Coulomb attraction or repulsion respectively. We are dealing with
Schr\" odinger equation
\be\label{Schr0} - \frac{\hbar^2}{2m}\,\Delta \psi({\bf
x})\,-\,\frac{q}{r} \psi({\bf x})\,=\,E \psi({\bf x}),\ \ r =
|{\bf x}| > 0 \ee
in the Hilbert space ${\cal H}_0$ specified by the norm
\be\label{norm0} \|\psi\|^2_0\,=\,\int\,d^3{\bf x}\ |\psi({\bf
x})|^2 \,. \ee
Expressing the wave function as
\be\label{psi0} \psi({\bf x})\,=\,R_j(r)\,H_{jm}({\bf x}),\ \
H_{jm}({\bf x}) \,\sim\,r^j\,Y_{jm}(\vartheta,\varphi)\,, \ee
and putting $\alpha = m q/\hbar^2$ and $k^2 = 2m E /\hbar^2$,
we obtain the radial Schr\" odinger equation:
\be\label{rSchr00}  r\,R^{\prime\prime}_j(r) + (2j+2) R^\prime_j(r)
+ 2 \alpha R_j(r)\,=\,-\,k^2 r\,R_j(r)\,. \ee
The parameter $\alpha$ is simply related to the H-atom Bohr radius $a_0 = \hbar^2/(m e^2) = |\alpha|^{-1} $. The solutions of (\ref{rSchr00}) are given in terms of solutions of the confluent hypergeometric equation (see, e.g. \cite{Schiff}).
%\be\label{rad0} R_j(r)\,=\,e^{-\kappa r}\,\phi\left(j+1-
%\frac{\alpha}{2\kappa},\,2j+2;\,2\kappa r\right)\,, \ee
%where $\phi(a,c;x)$ is the confluent hypergeometric function.

(i) For negative energies  one
obtains bound states with discrete energy eigenvalues:
\be\label{ener0} E_n\,=\,
-\frac{m\,e^4}{2\hbar^2 n^2}\ ,\ \ \ n\,=\,j+1,\,j+2,\,\dots\,. \ee

(ii) For positive $E\,=\,\frac{\hbar^2}{2m}\, k^2 >0$ one obtains
scattering states with $j$-th partial wave $S$-matrix
\be\label{scat} S_j (k)\ =\ \frac{\Gamma(j\,+\,1\,-\,
\frac{\alpha}{ik})}{\Gamma(j\,+\,1\,+\,\frac{\alpha}{ik})} \ .\ee
\\
This paper is organized as follows. We define the NC QM Coulomb problem in Section 2 : (i) We
define the rotationally invariant NC configuration space ${\bf R}^3_\lambda $ - the NC analog of
the Coulomb problem configuration space ${\bf R}^3_0$ and the Hilbert space ${\cal H}_\lambda $ of
wave functions in ${\bf R}^3_\lambda $, then, (ii) we introduce an important hermitian operator -
 the NC Coulomb problem Hamiltonian. In Section 3 we derive
the NC analog of the  radial Schr\"odinger equation, and we solve it for all energies. For energy $E<0$
we recover NC bound states regular in our non-commutativity parameter $\lambda $, while for $E>0$ one
obtains two sectors - low energy scattering regular in $\lambda $, and for ultrahigh energies
there are solutions singular in $\lambda$ which disappear in the commutative limit (this extends our
partial results for $E<0$ in \cite{GaP1}). The last Section 4 contains discussion and conclusions.

\section{The non-commutative space $\boldsymbol{R}^3_\lambda$}
 The name "non-commutative quantum mechanics" may seem to consist of more words than neccesary, since one cannot think of a quantum theory lacking certain non-vanishing commutators - the underlying uncertainty cannot be detached from the theory. The way from QM to NCQM can be roughly described as an analogy of the transition from the classical theory to QM. In the latter one the Heisenberg principle results into fuzziness of the phase space. In NCQM even the notion of a single point in the configuration space loses relevance. This fact is reflected in the non-vanishing commutator of the coordinates involved. Since we are about to deal with the Coulomb problem, the commutation relations have to preserve the rotational symmetry.

In this section we define the non-commutative (NC) space ${\bf
R}^3_\lambda $, possessing full rotational invariance, as a sequence
of fuzzy spheres introduced, in various contexts, in \cite{Ber}.
Different fuzzy spheres are related in such a way that at large
distances we recover space ${\bf R}^3_0$ with the usual flat geometry.
A similar construction of a 3D NC space, as a sequence
of fuzzy spheres, was proposed in \cite{Jab1}. However, various
fuzzy spheres are related to each other in a different manner (not leading
to the flat space ${\bf {R}}^3_0$ at large distances). \vskip0.2cm

{\it\ \textbf{The non-commutative configuration space}}
\\
We realize the NC coordinates in
${\bf R}^3_\lambda $ in terms of 2 pairs of boson annihilation and creation
operators $a_\alpha$, $a^\dagger_\alpha$, $\alpha\,=\,1,2$,
satisfying the following commutation relations, see \cite{GP}:
\be [a_\alpha,a^\dagger_\beta]\,=\,\delta_{\alpha\beta },\ \
[a_\alpha,a_\beta]\,=\,[a^\dagger_\alpha, a^\dagger_\beta]\,=\,0\, .\ee
They act in an auxiliary Fock space ${\cal F}$ spanned by normalized
vectors
\be\label{funct} |n_1,n_2\rangle\ =\ \frac{(a^\dagger_1)^{n_1}\,(a^\dagger_2)^{n_2}}{
\sqrt{n_1!\,n_2!}}\ |0\rangle\,. \ee
Here $|0\rangle\,\equiv\,|0,0\rangle$ denotes the normalized
vacuum state: $a_1\,|0\rangle\ =\ a_2\,|0\rangle\ =\ 0$. We shall use the
notation ${\cal F}_n\, =\, \{ |n_1,n_2\rangle\,|\ n_1+n_2 = n\}$.
\\
The noncommutative coordinates $x_j$, $j\,=\,1,2,3$, in the
space ${\bf R}^3_\lambda $ are given as
\be\label{sph3} \ x_j\ =\ \lambda\,a^+\,\sigma_j\,a\ \equiv\
\lambda\,\sigma^j_{\alpha\beta}\,a^\dagger_\alpha\,a_\beta,\
j\,=\,1,2,3\,,\ee
where $\lambda$ is a universal length parameter and $\sigma_j $ are Pauli matrices.  The operator that approximates
the NC analog of the Euclidean distance from the origin is
$r\,=\,\lambda\,(N + 1)$, $N = a^\dagger_\alpha a_\alpha$.
The coordinates $x_j$ and $r$ satisfy rotationally invariant relations:
\be\label{ncx} [x_i,x_j]\ =\ 2i\,\lambda\,\varepsilon_{ijk}\,x_k\,,
\ \ \ \ [x_i,r]\,=\,0\,,\ \ \ \ r^2 - x_j^2\,=\, \lambda^2\,.\ee
We will provide a strong argument supporting the exceptional role of
$r$ later. \vskip0.2cm

{\it\ \textbf{Hilbert space ${\cal H}_\lambda $ of NC wave functions}}
\\
Let us consider the linear space of normal ordered analytic functions containing the same number of
creation and annihilation operators:
\be\label{Hilb} \Psi\ =\ \sum\, C_{m_1 m_2 n_1 n_2}\,(a^\dagger_1
)^{m_1}\,(a^\dagger_2)^{m_2}\,(a_1)^{n_1}\,(a_2)^{n_2}\,,\ee
where the summation is finite over nonnegative integers satisfying
$m_1+m_2 \,=\,n_1+n_2$.  Here ${\cal H}_\lambda $ is our denotation of the Hilbert space of
functions (\ref{Hilb}) which possess finite weighted Hilbert-Schmidt norm
\be\label{whs1} \| \Psi \|^2\ =\ 4\pi\,\lambda^3\,\mbox{Tr}
[(N+1)\,\Psi^\dagger\,\Psi]\ =\ 4\pi\,
\lambda^2\,\mbox{Tr}[r\,\Psi^\dagger\,\Psi] \,.\ee
The rotationally invariant weight $w(r)\,=\,4\pi\,
\lambda^2\,r$ is determined by the requirement that a ball
in ${\bf R}^3_\lambda $ with radius $r$ should
possess a standard volume in the limit $r\,\to\,\infty$. The
projector $P_n$ on the subspace ${\cal F}_0
\oplus\,\dots\,\oplus {\cal F}_n$, corresponds to the
characteristic functions of a ball with the radius $r = \lambda (N+1)$.
Therefore, the volume of the ball  in question is
\be V_r\ =\ 4\pi\,\lambda^3\,\mbox{Tr}[(N+1)\,P_n]\ =\
4\pi\,\lambda^3\,\sum_{k=0}^n (k+1)^2\ =\
\frac{4\pi}{3}\,r^3\,+\,o(\frac{\lambda}{r})\,.\ee
Thus, the chosen weight $w(r)\,=\,4\pi\,\lambda^2\,r$
has the desired property. \vskip0.2cm

{\it\ \textbf{Orbital momentum in ${\cal H}_\lambda $}}
\\
In ${\cal H}_\lambda $ we define
orbital momentum operators, the generators of rotations $L_j$, $j\,=\,1,2,3$, as follows
\be\label{L0} L_j\,\Psi\ =\ \frac{1}{2}\,[a^+\,\sigma_j\,a,\Psi],\ \ j\,=\,1,2,3\,.\ee
They are hermitian (self-adjoint) operators in ${\cal H}_\lambda $ and obey the standard
commutation relations
\be [L_i,L_j] \Psi\,\equiv\,(L_i L_j\,-\,L_j L_i)
\Psi\,=\, i\,\varepsilon_{ijk}\, L_k \Psi\ .\ee
%With respect to the rotations (\ref{L0}) the doublet of
%annihilation (creation) operators transforms as spinor (conjugated
%spinor), whereas the triplet of NC coordinates as vector
%\be\label{spin2}
%\[ \hat{L}_j\,a_\alpha\,=\,-\,\frac{i}{2}\,\sigma^j_{\alpha\beta}
%,a_\beta,\ \ \hat{L}_j\,a^\dagger_\alpha\ =\,\frac{i}{2}\
%sigma^j_{\beta\alpha}\ a^\dagger_\beta\,,\ \ \]
%\ee \be\label{vek2} \ee
%\ \hat{L}_i\,x_j\ =\ i\,\varepsilon_{ijk}\,x_k\,.\]
% \vskip0.5cm
The standard eigenfunctions $\Psi_{jm}$, $j =
0,1,2,\,\dots,\,$, $m = -j,\,\dots,\,+j$, sa\-tisfying
\be\label{Hjm1} L^2_i\,\Psi_{jm}\ =\ j(j+1)\,\Psi_{jm},\ \ \
L_3\,\Psi_{jm}\ =\ m\,\Psi_{jm}\ ,\ee
are given by the formula
\be\label{harm2} \Psi_{jm}\ =\ \sum_{(jm)}\
\frac{(a^\dagger_1)^{m_1}\,(a^\dagger_2)^{m_2}}{m_1!\,m_2!}\ R_j(\varrho
)\ \frac{a^{n_1}_1\,(-a_2)^{n_2}}{n_1!\ n_2!}\,, \ee
where $\varrho = \lambda a^\dagger_\alpha a_\alpha = \lambda N$.
The summation goes over all nonnegative integers satisfying
$m_1+m_2 \,=\,n_1+n_2\,=\,j$, $m_1-m_2-n_1+n_2\,=\,2 m$. Thus
$\Psi_{jm} = 0$ when restricted to the subspaces ${\cal
F}_n$ with $n < j$ (what corresponds to the fact that in the standard QM the
first $j-1$ derivatives of $\Psi_{jm}$ vanish at the
origin). For any fixed $R_j(\varrho)$ equation
(\ref{harm2}) defines a representation space for a unitary irreducible
representation with spin $j$. \vskip0.2cm

{\it\ \textbf{The radial part and normalization in ${\cal H}_\lambda $}}
\\
%The symbol $:R_j(\varrho):$ represents a normal ordered form of the analytic function $R(\varrho)$:
%
%\be\label{harm3} :R_j(\varrho):\ =\ \sum_k c^j_k\,:
%\varrho^k:\ =\ \sum_k c^j_k \lambda^k\,\frac{N!}{(N-k)!}. \ee
%
%The last equality follows from the equation
%
%\be\label{nk} :N^k:\,|n_1,n_2\rangle\ =\ \frac{n!}{(n-k)!}\
%|n_1,n_2 \rangle,\ \ \ n\,=\,n_1 + n_2 \ee
%
%(which can be proved by induction in $k$). Since, $:N^k:\,
%|n_1,n_2\rangle\,=\,0$ for $k\,>\,n_1 + n_2$, the summation in
%(\ref{harm3}) is effectively restricted to $k \le n$ on any
%subspace ${\cal F}_n$.
%\section{Quantum mechanics in space $\boldsymbol{\hat{R}^3_0}$}
The two wave functions $\Psi_{jm}$ and $\tilde{\Psi}_{j'm'}$, with
$(j,m) \neq (j',m')$ and arbitrary factors $R_j(\varrho)$ and
$\tilde{R}_{j'}(\varrho)$, are orthogonal in ${\cal H}_\lambda $. Thus, when evaluating the norm of $\Psi_{jm}$, it is sufficient
to calculate  $\|\Psi_{jm}\|^2\,=\,\|\Psi_{jj}\|^2$ (this equality
follows from the rotational invariance of the norm in question):
\be\label{psi2} \|\Psi_{jm} \|^2\ =\ 4\pi \lambda^3\,\sum_{n=j}^\infty
\,\sum_{k=0}^n (n+1)\ \langle k,n-k|\,(n+1)\,\Psi_{jj}^\dagger
\,\Psi_{jj}\,|k,n-k\rangle\,.\ee
We benefit from the fact that $\Psi_{jj}$ has a simple form
\be\label{psijj} \Psi_{jj}\ =\ \frac{\lambda^j}{(j!)^2}\
(a^\dagger_1)^j \,R_j(\varrho)\,(-a_2)^j \,.\ee
The matrix element we need to calculate is
\[ \langle k,n-k|\ (a^\dagger_2)^j \,R_j(\varrho)\,a^j_1\
(a^\dagger_1)^j\,R_j(\varrho)\,a^j_2\ |k,n-k\rangle \]
\be\label{mel}  =\ \frac{(k+j)!(n-k)!}{k!\,(n-j-k)!}\
|{\cal R}_j(n-j)|^2\,, \ee
where
\be R_j(n)\,=\,\langle k,n-k|R_j(\varrho)
|k,n-k\rangle \ee
(the expression on the r.h.s. is $k$ - independent). Inserting (\ref{psijj}),
(\ref{mel}) into (\ref{psi2}) and using the identity (see \cite{Pr})
\[ \sum_{k=0}^{n-j}\ {{k+j}\choose j}\ {{n-k}\choose j}\ =\
{{n+j+1}\choose {2j+1}}\,,\]
we obtain
\be\label{Psi2} \|\Psi_{jm} \|^2\ =\ \frac{4\pi
\lambda^{3+2j}}{(j!)^2}\ \sum_{n=0}^\infty\ (n+j+1)\
{{n+2j+1}\choose {2j+1}}\ | R_j(n)|^2 \,. \ee
This expression represents, up to an eventual normalization, the
square of a norm of the radial part of the wave function.
%        Moreover, {\it the operator function
%          $R_j= \ :{\cal R}_j(\hat{\varrho}):$ possesses a finite norm in
%          ${\cal H}_\lambda $ provided the function
%          ${\cal R}_j\,=\,{\cal R}_j(\varrho)$ has finite norm in ${\cal
%         H}$} (since the norm (\ref{whs1}) asymptotically reduces to the
%         usual QM norm).

\skip0.2cm

{\it\ \textbf{The NC analog of Laplace operator in ${\cal H}_\lambda $}}
\\
We postulate the NC analog of the usual Laplace operator in the form:
\be\label{Lapl} \Delta_\lambda\,\Psi\ =\ -\,\frac{1}{\lambda r}\
[a^\dagger_\alpha,\,[a_\alpha,\,\Psi]]\ =\ -\,\frac{1}{\lambda^2
(N+1)}\ [a^\dagger_\alpha ,\,[a_\alpha ,\,\Psi]]\,.\ee
This choice is motivated by the following facts: (i) A double commutator
is an analog of a second order differential operator, (ii) the factor
$r^{-1}$ guarantees that the operator $\Delta_\lambda$ is hermitian
(self-adjoint) in ${\cal H}_\lambda $, and finally, (iii) the factors
$\lambda^{-1}$, or $\lambda^{-2}$ respectively, guarantee the correct
physical dimension of $\Delta_\lambda$ and its non-trivial commutative
limit.\vskip0.2cm
Calculating the action of (\ref{Lapl}) on  $\Psi_{jm}$ given in
(\ref{harm2}) we can check whether the postulate (\ref{Lapl}) is a
reasonable choice.\\
The operator $R_j(\varrho)$ in (\ref{harm2}) can be represented as a normal ordered expansion of an analytic function ${\cal R}_j(\varrho)$ :
\be\label{harm3} R_j(\varrho)\ = \,:{\cal R}_j(\varrho): \, = \sum_k c^j_k\,:
\varrho^k:\ =\ \sum_k c^j_k \lambda^k\,\frac{N!}{(N-k)!}. \ee
The last equality follows from the equation
\be\label{nk} :N^k:\,|n_1,n_2\rangle\ =\ \frac{n!}{(n-k)!}\
|n_1,n_2 \rangle,\ \ \ n\,=\,n_1 + n_2 \ee
(which can be proved by induction in $k$). Since $:N^k:\,
|n_1,n_2\rangle\,=\,0$ for $k\,>\,n_1 + n_2$, the summation in
(\ref{harm3}) is effectively restricted to $k \le n$ on any
subspace ${\cal F}_n$.\\
\\
The following formula follows from commutation relations (\ref{A2 }); the proof is given in Appendix A:
\[ [a^\dagger_\alpha ,\,[a_\alpha,\,\Psi]]\ =\ \lambda^j\ \sum_{(jm)}\
\frac{(a^\dagger_1)^{m_1}\,(a^\dagger_2)^{m_2}}{m_1!\, m_2!}\times \]
%\frac{\lambda^j}{2\lambda \hat{r}}\
\be\label{rrov0} \times\ \ :[ - \varrho\,{\cal R}''(\varrho )\,-\,2(j+1)\,
{\cal R}'(\varrho )]:\
\frac{a^{n_1}_1\,(-a_2)^{n_2} }{n_1!\ n_2!}\,.\ee
Here ${\cal R}'(\varrho )$ denotes the usual derivative
 defined as:
\be\label{derivative0} {\cal R}(\rho )\,=\,\sum_{k=0}^\infty\,c^j_k\,\rho^k\ \ \ \Rightarrow
\ \ \  {\cal R}'(\varrho)\,=\,\sum_{k=1}^\infty\, k\,c^j_k\,\varrho^{k-1}\,,\ee
and ${\cal R}''(\varrho )$ is defined as the derivative of ${\cal R}'(\varrho )$.
Thus, the prime corresponds exactly to the usual derivative
$\partial_\varrho$. In the commutative limit
$\lambda\,\rightarrow\,0$ operator $\varrho$ formally reduces to the usual
radial $r$ variable in ${\bf R}^3$, and we see that $\Delta_\lambda$ just
reduces  to the standard Laplace operator in ${\bf {R}}^3$. \vskip0.2cm

{\it\ \textbf{The potential term in  ${\cal H}_\lambda $}}
\\
The operator $V$ corresponding to a central potential in QM is defined simply as the
multiplication of the NC wave function by $V(r)$:
\be (V \Psi)(r)\ =\ V(r)\,\Psi \ =\ \Psi\,V(r)\,.\ee
Since any term of $\Psi\,\in\,{\cal H}_\lambda $ contains the same number of
creation and annihilation operators (any commutator of such a term with $r$ is zero), the left and right multiplications by
$V(r)$ are equal.

In the commutative case the Coulomb potential is a radial solution
of the equation (\ref{coul0}) vanishing at infinity. Due to our
choice of the NC Laplace operator $\Delta_\lambda$ the
 NC analog of this equation is
\[  \Delta_\lambda\,V(r)\ =\ 0\ \ \ \Leftrightarrow \ \ \
[\,a^\dagger_\alpha,\, [\,a_\alpha\,,V(N)\,]\,]\ =\ 0\,. \]
Last equation can be rewritten as a simple recurrent relation
\be\label{rec} (N+2)\,V(N+1)\,-\,(N+1)\,V(N)\ =\
(N+1)\,V(N)\,-\,N\,V(N-1)\,.\ee
Putting $(M+1)\,V(M)\,-\,M\,V(M-1)\,=\,q_0$ and $V(0)\,=\,q_0\,-\,\frac{q}{\lambda}$,
and summing up the first equation over $M\,=\,1,\,\dots\,N$, we
obtain the general solution:
\be\label{nccoul1} V(N)\ =\ -\,\frac{q}{\lambda\,
(N+1)}\ +\ q_0\ =\ -\,\frac{q}{r}\ +\ q_0\,,\ee
where $q$ and $q_0$ are arbitrary constants ($\lambda$ is
introduced for the future convenience). Thus the NC analog of the Coulomb potential
vanishing at infinity is given by
\be\label{nccoul}(V \Psi)(r)\ =\ -\frac{q}{r}\,\Psi\,. \ee
We see that the $\frac{1}{r} = \frac{1}{\lambda (N+1)}$ dependence  of the NC
Coulomb potential is inevitable.

\section{The Coulomb problem in NC QM}
Based on (\ref{Lapl}) and (\ref{nccoul}) we postulate the NC analog of the
Schr\"odinger equation with the Coulomb potential in ${\bf R}^3_\lambda$ as
\be\label{ncsch1} \frac{\hbar^2}{2m\lambda r}\,[a^\dagger_\alpha ,
[a_\alpha ,\Psi]] - \frac{q}{r}\, \Psi = E\,\Psi\ \ \
\Leftrightarrow\ \ \ \frac{1}{\lambda}\,[a^\dagger_\alpha,
[a_\alpha ,\Psi]] - 2\alpha\,\Psi = k^2\,r \Psi\,,\ee
%\,,\ee
\newline
In Appendix A these two equations are proved:
\[ [a^\dagger_\alpha ,\,[a_\alpha ,\,\Psi_{jm}]]\ =\ \sum_{(jm)}\ \dots\
:[ -\varrho \lambda \,{\cal R}_j^{\prime\prime}\,-\,2(j+1) \lambda \,{\cal R}_j^\prime]:\ \dots \ ,\]
\be\label{appen} r\,\Psi_{jm}\ =\ \sum_{(jm)}\ \dots\
:[(\varrho+\lambda j+\lambda )\,{\cal R}_j\,+\,\lambda\,\varrho\,{\cal R}_j^\prime]:
\ \dots \ ,\ee
where ${\cal R}_j \equiv {\cal R}_j(\varrho )$ and similarly for derivatives. The dots on
the left and right in (\ref{appen}) denote the products in $\Psi$ containing
respectively creation and annihilation operators together with the factor
$\lambda^j$, that represent the angular dependence of $\Psi$
and remain untouched as the operators in question are rotation
invariant. Inserting (\ref{appen}) into (\ref{ncsch1}) we obtain the NC
analog of radial Schr\"odinger equation:
\be\label{radNC} : \varrho\,{\cal R}_j^{\prime\prime}
\,+\,[k^2 \lambda \varrho + 2j+2]\,{\cal R}_j^\prime\,+\,[k^2 \varrho\,+
\,k^2 \lambda (j+1)\,+\,2\alpha]\,{\cal R}_j:\ =\ 0\,.\ee

We claim (\ref{radNC})  to be an NCQM analog of the usual radial Schr\"{o}dinger equation (\ref{rSchr00}) known from QM. There definitely is a  resemblance; in the limit $\lambda \rightarrow 0$ the terms in (\ref{radNC}) proportional to $\lambda $  representing the NC corrections disappear. Considering the same limit we see that the presence of the colon marks denoting the normal ordering should not worry us either; recall that for zero $\lambda$ it makes no difference whatsoever whether we care for the ordering   or not. (Normal and usual powers coincide for zero $ \lambda$.) This is a good news to start with, leaving us, however, with the task to solve (\ref{radNC}) for nonzero $\lambda$, which means that both the extra terms proportional to $\lambda$ and the normal ordering are to be taken at a face value. If it was not for the ordering issues, the solution would be quite straightforward - the extra terms would mean just adding some more work needed to complete the calculation, but it is known how to solve the problems of this kind. In fact this is precisely what we are going to do:
We associate the following ordinary differential equation to the
mentioned operator radial Schr\"odinger equation (\ref{radNC}):
\be\label{rad} \varrho\,{\cal R}_j^{\prime\prime}
\,+\,[k^2 \lambda \varrho + 2j+2]\,{\cal R}_j^\prime\,+\,[k^2 \varrho\,+
\,k^2 \lambda (j+1)\,+\,2\alpha]\,{\cal R}_j\ =\ 0\,,\ee
with $\varrho$ being real variable, and we will solve this one. But how come we expect this step to be of any use to us, when we actually \textit{do} have to care about the ordering ? One should notice the following: whatever appears in the equations, it can be expressed in terms of powers in $\varrho$ ; it is just that they are normal powers in one case and the usual powers in the other. Next, we have some operators in both equations; normal derivatives in one case and the usual ones in the other. The key information is, that the derivative defined in (\ref{derivative0}) acts on the normal powers just like a carbon copy of usual  derivative, see (\ref{A2 }).
\\
Now bearing this in mind, we expect $R \, =\, :{\cal R}:$ , the solution of (\ref{radNC}), to be of the same form as $\cal R $, the solution of (\ref{rad}), except for the nature of the powers involved. So a brief summary goes like this: The solution of (\ref{rad})with all the usual powers replaced by the normal ones is the solution of (\ref{radNC}).  However, the form of the solution is not the best one yet. We have already mentioned the relation between the equation given by QM (\ref{rSchr00}) and (\ref{radNC}), the one supplied by NCQM. Of course we would like to compare the corresponding solution as well, but this is rather a difficult task as long as we have normal powers in the first  and the usual ones in the latter one. Fortunately we have a formula relating $:\varrho^n:$ and $\varrho^n$, namely
\be :\varrho^n : \, = \lambda^n : N^{n}: = \lambda^n \frac{N!}{(N-n)!}\ ,\ \ \ \,:\varrho^{-n} : \, = \lambda^{-n} : N^{n}: = \lambda^{-n}\frac{N!}{(N-n)!} \,  \ee
All we need is to rewrite $:{\cal R} :$ using those relations. Then the above mentioned comparison of QM and NCQM will be obtained.
\\
\\
Perhaps is has been made clear enough what is to be done, so let's get started with the solution of equations (\ref{rad}) and (\ref{radNC}). Some mathematic theory is to be studied here.  We have got a second order differential equation of the form
\be\label{general}
\left( a_0 x+ b_0\right)y''(x) +\left( a_1x+ b_1\right)y'(x) +\left( a_2x+ b_2\right)y(x) =0
\ee
Depending on whether the quantity $ D^2 \equiv a_1^2 -4 a_0 a_2 $ is zero or not, the solutions of (\ref{general}) are given in terms of Bessel or confluent hypergeometric functions respectively. We will restrict ourselves to the case $a_0 = 1, \,\, b_0 =0 $ at the price of some generality loss -but generality is not what we are after in the first place.
\\
\\
\textbf{(i) $a_0 = 1$ , \,\, $b_0=0$, \,\,  $ D^2 =a_1^2 -4 a_0 a_2 \neq 0$} \vskip0.2cm
In this case, the solution of (\ref{general}) is of the form
\be\label{form} y(x) \ = \ e^{\frac{D-a_1}{2}} \textit{Z}(a, c, -Dx) ,\ee
where $ a=\frac{1}{D}\left( \frac{D-a_1}{2}b_1 +b_2    \right)\mbox{,\,\,}c=b_1 ,$ and  $\textit{Z} (a, c,\tilde{x})$  is any solution of the confluent hypergeometric equation
\be\label{confl} \tilde{x}\,\textit{Z}''\,+\,(c-\tilde{x} )\textit{Z}'
\,-\,a\,\textit{Z}\ =\ 0\,. \ee
The solution of (\ref{confl}) regular at the origin will be the most important to us. It is often referred to as the confluent hypergeometric function:
\be\label{y1}  \phi (a;c;\tilde{x} )\ =\ \sum^\infty_{m=0}\ \frac{(a)_m }{(c)_m}\, \frac{{\tilde{x}}^m}{m!}\,, \ee
where $(a)_m$ denotes the so-called Pochhammer symbol: $(a)_0\,=\,1 $ and
$$ (a)_m = a (a+1)\ \dots\ (a+m-1)\,=\,\frac{\Gamma(a+m)}{\Gamma(a)}\,, \ \ \ \ m\ =\ 0, 1,2,\dots \ . $$
The fundamental system of solutions of (\ref{confl}) consists of the following functions:
\be\label{y5-7} \psi(a,c;\tilde{x} )\,,\ \ \ \mbox{and\,} \ \ \ e^{\tilde{x} }\psi(c-a,c;-\tilde{x} )\,,\ee
with $\psi(a,c;\tilde{x} )$ possessing this asymptotic expansion for $\tilde{x} \,\to\,\infty$ \, :
\be\label{psi} \psi (a,c;\tilde{x} )\ =\ \sum^\infty_{m=0} (-1)^m\frac{(a)_m (a-c+1)_m} {m!}\, {\tilde{x}}^{-a-m}\,. \ee
Every solution (e.g. also (\ref{y1}), the one regular at the origin) can be expressed as a suitable linear
combination of (\ref{y5-7}). This possibility is useful when treating the scattering processes .\\
\\
Note that $D$ is fixed up to the sign, since the coefficients in the equation determine the value of $D^2$ only. In fact it does not matter which possibility is preferred in (\ref{form}). Replacing $D$ with $-D$  makes no difference  because of the Kummer identity which reads
\be\label{Kid} e^{-\tilde{x}/2}\,\phi (a,c;\tilde{x})\, =\, e^{\tilde{x}/2}\,\phi (a,c; -\tilde{x})\,.\ee
\newline
\textbf{(ii) $a_0 = 1$ , \,\, $b_0=0$, \,\,  $ D^2 =  a_1^2 -4 a_0 a_2 =0$ }\vskip0.2cm
This time the solution of (\ref{general}) has the following  form:
\be\label{formbessel} y(x)\ = \ e^{-\frac{a_1}{2}x}        x^{\frac{1-b_1}{2}}  C_{1-b_1} \left( \sqrt{(-2a_1b_1 +4b_2)x}\right) \ee
$C_\nu  (\tilde{x})$ is any solution of the Bessel equation
\be\label{bessel} \tilde{x}^2 C_\nu  ''+\tilde{x}C_\nu  '  +(\tilde{x}^2 -\nu^2)C_\nu =0 \ee
We require regularity in the origin, what leads us to the Bessel function:
\be\label{form bessel} C_\nu  (\tilde{x})=J_{\nu}(\tilde{x})=\sum^\infty_{m=0}(-1)^m \left( \frac{\tilde{x}}{2}\right)^{2m +\nu}\frac{1}{m!\Gamma(m+\nu+1)} \ee
After this general mathematic interlude we return to our equation (\ref{rad}). For the sake of brevity it is suitable to introduce a new parameter $\eta$ defined as
\be \eta=\frac{k \lambda}{2}=\frac{\sqrt{2mE}\lambda}{2\hbar} \ee
According to (\ref{rad}) the following holds:
\be\label{coef}
\begin{array}{lll}
a_0 =1 & \,\,\,\,\, a_1 =\lambda k^2 =\frac{4\eta^2}{\lambda}& \,\, a_2 =k^2=\frac{4\eta^2}{\lambda^2} \\
%& & \\
b_0 =0 & \,\,\,\,\, b_1 =2(j+1) & \,\, b_2 =\lambda k^2  (j+1)+\alpha =\frac{4}{\lambda}(j+1)+2\alpha\\
%& & \\

\end{array}
\ee
$$ \Rightarrow \ \ \ \ D =\pm(\lambda^2 k^4 -4 k^2 )^{1/2} = \pm \frac{4}{\lambda}\sqrt{\eta^2\left(  \eta^2 -1 \right)}
$$
For $D\neq 0$, or equivalently $\eta \neq 0, \, \eta\neq 1$ the solution of (\ref{radNC}) is
\be\label{reg3}
\begin{array}{lll}
R_{j\pm} & =& \, :{\cal R}_{j \pm}:\\
          & = & \, :\exp\left[\left(\pm\frac{2\eta\sqrt{\eta^2-1}}{\lambda}-\frac{2\eta^2}{\lambda}\right)\varrho\right] \times \\
   & & \times  \phi\left(j+1\pm\frac{\alpha\lambda}{2\eta\sqrt{\eta^2-1}}, \, 2j+2, \, \mp\frac{4\eta\sqrt{\eta^2-1}}{\lambda}\right):\\
   &=& ...\mbox{see Appendix}...\\
   & =& \left[1\pm 2\eta\sqrt{\eta^2-1}-2\eta^2\right]^N \times \\
   & & \times  F\left(j+1\pm\frac{\alpha\lambda}{2\eta\sqrt{\eta^2-1}},\,-N ,\, 2j+2, \, \pm\frac{4\eta\sqrt{\eta^2-1}}{1\pm 2\eta\sqrt{\eta^2-1}-2\eta^2}\right)\\
\end{array}
\ee
\\
$F (a,b;c;z)$ is the  usual hypergeometric function:
\be\label{hypgeo} F (a,b;c;z)=\sum^\infty_{m=0} \frac{(a)_m(b)_m}{(c)_m} \frac{z^m}{m!}, \ee
It is one of the solutions of the hypergeometric equation
\be\label{hypgeoeq} z (1-z)u''(z) + [c - (a+b+1)z]u'(z) - ab u(z)=0 . \ee
The calculations needed to get rid of the normal ordering  in the above equation are briefly sketched in Appendix.
The $\pm $ signs that emerged as a lower index in $R_{\pm}$  spring from the two possible choices of the sign of $D$. We have mentioned that the choice of sign is completely arbitrary due to the Kummer identity which holds for the confluent hypergeomatric function. This fact survives the process of rewriting the normal powers in terms of the usual ones and is reflected in the so-called Euler identity for the hypergeometric functions:
\be\label{Eid} F (a,b,c;x)\, =(1-x)^{-b} F(c-a,\, b,\, c;\,x(x-1)^{-1} )\,.\ee
\\
If $\eta=0$, the solution of (\ref{radNC}) is
\be\label{gsol2}
\begin{array}{lll}
R_{j} & =& \, :{\cal R}_{j}:\\
 & =& :\varrho^{-j-1/2}\,J_{-2j-1}(\sqrt{8\alpha\varrho}): \\
   &= & ...\mbox{see Appendix} ... \\
   &=& -\frac{(2\alpha)^{j+1/2}}{(2j+1)!}\, \phi(-N,\, 2j+2,\, 2\alpha\lambda)
\end{array}
\ee
\\
And finally for $\eta=1$
\be\label{gsol3}
\begin{array}{lll}
R_{j} & =& \, :{\cal R}_{j }:\\
& =& :e^{2\varrho/\lambda}\varrho^{-j-1/2}\,J_{-2j-1}(\sqrt{8\alpha\varrho}): \\
   &= & ...\mbox{see Appendix}... \\
   &=& -(-1)^N\frac{(2\alpha)^{j+1/2}}{(2j+1)!}\, \phi(-N,\, 2j+2,\, -2\alpha\lambda)
\end{array}
\ee
\\
To sum up, the solution of (\ref{radNC}) typically consists of an exponential factor (or its NCQM analog,  see Appendix for further explanation) multiplied by certain power series. Depending on the energy, the exponent may be real or imaginary.
We are supposed to investigate  the following cases:\\
\\
(i) $\eta \in i \textbf{R} \, \, \,\mbox{or} \,\,E<0$;\\
(ii)  $\eta \in (1, + \infty) \,  \, \, \mbox{or}\,\, E>\frac{2\hbar^2}{m\lambda^2}$;\\
(iii)  $\eta \in (0, 1 )  \, \, \, \mbox{or} \,\, 0<E< \frac{2\hbar^2}{m\lambda^2}$;\\
(iv)  $\eta=0 \,\, \,\mbox{or} \,\, E=0$;\\
(v)  $\eta =1 \,\, \, \mbox{or}\,\,  E= \frac{2\hbar^2}{m\lambda^2} $ \\
\\
In (i) and (ii)  the arguments in the exponentials are real - bound states may occur for certain energy values. The (iii) leads to the scattering, the exponential factor involves an imaginary part. The last two, (iv) and (v), are the  border cases separating the above mentioned intervals.

\subsection{Bound states}
At the very beginning it is suitable to remind us of the QM version, which predicts bound states to occur under the condition that the potential is attractive $(\alpha > 0)$ and the energy has some specific negative values mentioned in (\ref{ener0}), i.e.
$$E_n\,=\, -\frac{m\,e^4}{2\hbar^2 n^2}\, ,\, \, \, n\,=\,j+1,\,j+2,\,\dots\,. $$
Now to NCQM. In this section, the $R_+$ form of the solution of (\ref{radNC}) will be suitable for our purposes. The reason is, that in $R_+$ the absolute value of the factor multiplying the hypergeometric function is less than 1 for every $N$. Consequently, when looking for bound states, is is sufficient to check whether the power series terminates. In case of hypergeometric function this happens if the first argument is a negative integer. This leads to discrete energy values.
\skip0.2cm

\subsubsection{\it\ \textbf{Bound states for $E<0$ , \, $\eta = i \, |\eta|$}}
The equation (\ref{reg3}) in this case reads
\begin{eqnarray}\label{solbound1}
 R_j  &=& \left[1 - 2|\eta|   \sqrt{|\eta|^2+1}  +2|\eta|^2 \right]^N\times     \nonumber \\
   & &  \times F\left( j+1- \frac{\alpha\lambda}{2|\eta|\sqrt{|\eta|^2+1}} , -N; 2j+2; \frac{- 4|\eta|\sqrt{|\eta|^2+1}}{1 - 2|\eta| \sqrt{|\eta|^2+1}  +2|\eta|^2 }\right)\nonumber \\
   & &
\end{eqnarray}
Square integrable bound state solutions can be easily seen, since the hypergeometric function has to reduce to a  polynomial. This happens only if $\alpha > 0$, i.e. in the {\it Coulomb attractive} case, provided that
\be\label{BS}  \alpha > 0\ \ \ \ \ \mbox{and}\ \ \ \ \ j+1 - \frac{\alpha\lambda}{2|\eta|\sqrt{|\eta|^2+1}}\, =\,-n\,.\ee
This gives the bound state energies (with Planck constant $\hbar$ and mass $m$ explicitly introduced):
\be\label{energy}
E^{I}_{\lambda \,n} \ =\ \frac{me^4}{2\hbar^2 n^2}\, \frac{2}{1+ \sqrt{1+(\alpha \lambda/n)^2}}\ =\
\frac{\hbar^2}{m\lambda^2}\left(1- \sqrt{1+(\alpha\lambda/n)^2}\right)\,. \ee
Taking the limit $\lambda \to 0$ we recover the QM result. We see that the NC corrections are governed by $(\lambda \alpha)^2 = (\lambda/a_o)^2 $ - the square of the ratio of two dimension-full parameters in the model - the noncommutativity parameter $\lambda $ and the Bohr radius $a_o$.

\skip0.2cm

\subsubsection{\it\ \textbf{Bound states for $E>2/\lambda^2$ , \, $\eta = |\eta|>1$}}
The equation (\ref{reg3}) in this case is
\begin{eqnarray}\label{solbound2}
 R_j  &=& \left[1 + 2|\eta|   \sqrt{|\eta|^2-1}  -2|\eta|^2 \right]^N\times     \nonumber \\
   & &  \times F\left( j+1+\frac{\alpha\lambda}{2|\eta|\sqrt{|\eta|^2-1}} , -N; 2j+2; \frac{4|\eta|\sqrt{|\eta|^2-1}}{1 + 2|\eta| \sqrt{|\eta|^2-1}  -2|\eta|^2 }\right)\nonumber \\
   & &
\end{eqnarray}

Since the absolute value of the prefactor preceding the hypergeometric function is less than 1, the whole solution can have a finite norm. There is a possibility for  the hypergeometric function to terminate, since the first argument becomes a negative integer for certain energy values, under the condition that $\alpha<0$ (so  the potential has to be repulsive this time).
\be  \alpha <0\ \ \ \ \ \mbox{and}\ \ \ \ \ j+1+\frac{\alpha}{k\sqrt{|\eta|^2-1}} =\,-n\,.\ee
In this case the bound state energies read:
\be\label{energy1}
E^{II}_{\lambda \, n} \ =\ \frac{\hbar^2}{m\lambda^2}\left(1 + \sqrt{1+(\alpha\lambda/n)^2}\right)\
=\ \frac{2\hbar^2}{m\lambda^2} - E^{I}_{\lambda \, n} . \ee
These are very unexpected solutions, $E^{II}_{\lambda \,n}$ being a mirror of $E^I_{\lambda \, n}$ with respect to the "critical energy" $E_{crit} = 2\hbar^2/(m\lambda^2)$. However, they disappear from the Hilbert space ${\cal H}_\lambda $ in the commutative limit $\lambda \to 0$.\\
For energies $E=E^{II}_{\lambda \,n}$ the solution (\ref{solbound2}) has the same finite norm as (\ref{solbound1}) for $E=E^{I}_{\lambda \,n}$.
\vskip0.2cm

\subsection{Scattering  $E\in(0, \, 2/\lambda^2)$, \, $\eta \in (0,1)$ }

In this section we will deal with the NCQM version of Coulomb scattering.  We will briefly sum up the QM results  before handling our NCQM case.
\\
The solution of the radial Schr\"{o}dinger equation for a particle in a potential  $V(r) = -\alpha /r$ with the angular momentum $j$ and energy $E > 0$, regular in  $r\to 0$ is given in terms of the confluent hypergeometric function  (see \cite{Schiff}):
\be\label{schiff1} R^{QM}_j\ =\ e^{ikr}\,\phi\left(j+1-i\frac{\alpha }{k},\,2j+2,\,-2ikr \right)\,,\ \ \ k\, =\, \sqrt{2E}\,>\,0\,.\ee
We have refrained from writing down $m/\hbar^2$ explicitly here. This will simplify the formulas and will not do any harm, since the full form can be restored anytime.\\
The solution is real and for  $r\rightarrow  \infty$ it can be written as a sum of two complex conjugated parts corresponding to an in- and out- going spherical wave. In the following formula a real term common for both parts is left out, having no influence on the S-matrix.
% e^{-\frac{\pi \alpha }{2k}}\ \frac{\Gamma (2j+2)}{(2k)^l} \
\be\label{schiff2} R^{QM}_j\ \sim \ \frac{i^{j+1}}{\Gamma (j+1 +i\frac{\alpha}{k})}\,e^{ikr +i\frac{\alpha }{k} \ln(2kr)}\ +\ \frac{i^{-j-1}}{\Gamma (j+1 -i\frac{\alpha}{k})}\,e^{-ikr -i\frac{\alpha }{k} \ln(2kr)} \,.\ee
The $S$-matrix for the $j$-th partial wave is defined as the ratio of the  $r$-independent factors multiplying the exponentials with the kinematical factor $(-1)^{j+1}$ left out.
\be\label{schiff3} S^{QM}_j(E)\ =\ \frac{\Gamma (j+1 -i\frac{\alpha}{k})}{\Gamma (j+1 +i\frac{\alpha}{k})}\,,\ \ \ \ E = \frac{1}{2}\,k^2 > 0 \,.\ee

Now let us have a look on the Coulomb scattering in NCQM, considering the  $j$-th partial wave and the energy $E\,\in\,(0,\,2/\lambda ^2)$. For the future convenience it is suitable to introduce
\be\label{NCries2}  p\, =\, \sqrt{2E (1 -\frac{1}{2}\lambda ^2 E) }\ .\ee
The solution regular in the origin is given in terms of the hypergeometric function (\ref{hypgeo}):
\be\label{NCries1} R_{E j}\ =\ \left( \frac{p+i\lambda E}{p-i\lambda E}\right)^N\,F\left(j+1-i\frac{\alpha }{p},\,-N,\,2j+2;\, 2i\lambda p\, \frac{p-i\lambda E}{p+i\lambda E}\right)\,\ee
We choose the positive  square root (\ref{NCries2}) for $E\,\in\,(0,\,2/\lambda ^2)$. The formula (\ref{NCries2}) represents a conformal map from the upper complex $E$-plane on a right complex  $p$-plane with a branch cut $p\,\in\,(0,\,1/\lambda )$.
The radial  dependence of  $R_j$ is present in the hermitian operator  $N$: $r = \varrho +\lambda $, $\varrho  = \lambda N$. In analogy with   (\ref{schiff2} ) we will rewrite also the NC solution as a sum of two terms corresponding to the in- and out- going spherical wave. Again, leaving out the common hermitian factor which is irrelevant regarding the $S$-matrix, we can write (see Appendix):

\beqa\nn R_{Ej} &\sim &  \frac{(-1)^{j+1}e^{\alpha\pi/p}}{\Gamma (j+1 +i\frac{\alpha}{p})}\,\left( \frac{p+i\lambda E}{p-i\lambda E}\right)^{N+j+1-i\frac{\alpha}{p}}\,\frac{\Gamma(2j+2)\Gamma(N+1)}{\Gamma(N+2+j-i\alpha /p)}\,(2\lambda p)^{-1-j+i\frac{\alpha}{p}}\\
&\times & F\left(j+1-i\frac{\alpha }{p},\,-j-i\frac{\alpha }{p},\,N+2+j-i\frac{\alpha }{p};\,\frac{-i}{2\lambda p}\, \left(\frac{p+i\lambda E}{p-i\lambda E}\right)\right) \nn\\
&+ &  \frac{(-1)^{j}e^{\alpha\pi/p}}{\Gamma (j+1 -i\frac{\alpha}{p})}\,\left( \frac{p-i\lambda E}{p+i\lambda E}\right)^{N+j+1+i\frac{\alpha}{p}}\,\frac{\Gamma(2j+2)\Gamma(N+1)}{\Gamma(N+2+j+i\alpha /p)}\,(2\lambda p)^{-1-j-i\frac{\alpha}{p}} \nn \\
&\times & F\left(j+1+i\frac{\alpha }{p},\,-j+i\frac{\alpha }{p},\,N+2+j+i\frac{\alpha }{p};\,\frac{i}{2\lambda p}\, \left(\frac{p-i\lambda E}{p+i\lambda E}\right)\right)\,. \label{NCries3} \eeqa

To enable better comparision with (\ref{schiff2}) let us  rewrite also (\ref{NCries3}) as a sum of two complex conjugated parts. Some sort of sketch of the calculation leading to it is to be found in the Appendix C:
\beqa\label{rozklad} R_{Ej} &\sim &  (-1)^{j+1}i^{-j-1}e^{-\alpha\pi/2p}\ \frac{\Gamma(2j+2)}{\Gamma (j+1 -i\alpha/ p)} \frac{e^{-i\frac{\alpha}{p}\ln(2pr)}}{(2pr)^{j+1}}\nn \\
&\times & \exp \left[-\left(r / \lambda+j+i \alpha/ p \right)\, \ln \frac{p+i\lambda E}{p-i\lambda E}\,\right] \nn \\
&\times & \exp \left[- \sum_{n=1}^\infty (\lambda / r)^n\, \frac{B_{n+1}(j+1+i\alpha /p)\,-\,B_{n+1}(0)}{n(n+1)} \right] \nn \\
&\times & F\left(j+1+i\frac{\alpha}{p},\,\,\,-j+i\frac{\alpha}{p},\,\,\,\frac{r}{\lambda} +j+1+i\frac{\alpha}{p};\,\,\,-\frac{1}{2i\lambda p}\, \frac{p-i\lambda E}{p+i\lambda E}\right) +\nn \\
& & \nn \\
& + &  (-1)^{j+1}i^{j+1}e^{-\alpha\pi/2p}\ \frac{\Gamma(2j+2)}{\Gamma (j+1+i\alpha/ p)} \frac{e^{i\frac{\alpha}{p}\ln(2pr)}}{(2pr)^{j+1}}\nn \\
&\times & \exp \left[-\left(r/ \lambda +j - i \alpha/ p \right)\, \ln \frac{p-i\lambda E}{p+i\lambda E}\,\right] \nn \\
&\times & \exp \left[- \sum_{n=1}^\infty (\lambda / r)^n\, \frac{B_{n+1}(j+1-i\alpha /p)\,-\,B_{n+1}(0)}{n(n+1)} \right] \nn \\
&\times & F\left(j+1-i\frac{\alpha}{p},\,\,\,-j-i\frac{\alpha}{p} ,\,\,\,\frac{r}{\lambda}+j+1-i\frac{\alpha}{p};\,\,\,\frac{1}{2i\lambda p}\, \frac{p+i\lambda E}{p-i\lambda E}\right) \nn \\
& &  \label{NCries4}\eeqa
%
%For $N \gg 1$ the first two lines in (\ref{NCries4}) are related to  (\ref{schiff2}), in fact they coincide for $\lambda \rightarrow 0$. The second and the third line in (\ref{NCries4}) are the NC corrections proportional to $1\,+\,O(\lambda /r)$ for $N \gg 1$ and do not enter the  $S$-matrix.
The $S$-matrix  is the ratio of the  $r$-independent factors :
\be\label{NCries5} S^\lambda_j(E)\ =\ \frac{\Gamma (j+1 -i\frac{\alpha}{p})}{\Gamma (j+1 +i\frac{\alpha}{p})}\ ,\ee
where
\be\label{NCries6} E\ =\ \frac{1}{\lambda^2}\, \left(1\,+\,i\,\sqrt{\lambda^2 p^2 - 1} \right) \ee
is the  conformal map inverse to (\ref{NCries2}), which maps the  cut $p$ right-half-plane into the  $E$ upper-half-plane. We take the positive square root in  (\ref{NCries6}) for $p\,\in\,(1/\lambda ,\,+\infty)$.
\\
The physical-relevant values of the $S$-matrix are obtained as $S^\lambda_j(E+i\varepsilon )$ in the limit $\varepsilon\,\rightarrow \,0_+ $.
\\
The interval corresponding to the scattering $E\,\in\,(0,2/\lambda ^2)$ is mapped onto the branch cut in the $p$-plane as follows:
\beqa\nn E \in (0,1/\lambda ^2)\ &\mapsto &  \mbox{upper edge of the branch cut} \ p\in\,(0,\,1/\lambda )\\
E \in (1/\lambda ^2,2/\lambda ^2)\ &\mapsto &  \mbox{lower edge of the  branch cut} \ p\in\,(0,\,1/\lambda ) \label{NCries7}
\eeqa

\subsection{Bound states revisited - poles of the S-matrix }
Like in the previous section, we will briefly remind the QM case: In the case of an attractive potential ($\alpha > 0$)  the $S$-matrix (\ref{schiff3}) has poles in the upper complex  $k$-plane for
\be\label{schiff4} k_n\ =\ i\,\frac{\alpha }{n}\,, \ \ \ n\, =\, j+1,\ j+2,\ \dots \ee
The wave function (\ref{schiff1}) is integrable  for $k\,=\,k_n$:
\be  R^{QM}_{nj}\ =\ e^{-\alpha \frac{r}{n}}\,\phi\left(n,\,2j+2,\,2\alpha \frac{r}{n} \right)\,.\ee
It is obvious that the energy levels correspond to the poles of the $S$-matrix:
\be\label{schiff5}  E_n\ =\ - \frac{\alpha^2}{2n^2}\,, \ \ \ n\, =\, j+1,\ j+2,\ \dots\ee
In NCQM there is an analogy,  the poles of the S-matrix  occur in the case of attractive potential $(\alpha>0)$ for some special values of energy below $0$. However, poles can be found also in case of repulsive potential $(\alpha<0)$ for particular values of energy above $2/\lambda^2$ ).

\subsubsection{\it\ \textbf{ Poles of the $S$-matrix for attractive potential}}
\[ p^\lambda_n\,=\,i\,\frac{\alpha }{n}\,,\ \ \ \alpha > 0\ \ \ \Leftrightarrow\ \ \ E^{I}_{\lambda \, n}\,=\,\frac{1}{\lambda^2}\, \left(1\,-\,\sqrt{1+(\lambda\alpha/n )^2 } \right)\, < 0.\]
\be\label{NCries8}  n\, =\, j+1,\ j+2,\ \dots \ee
In the limit $\lambda \rightarrow 0$ this coincides with the standard self-energies of the hydrogen atom (\ref{schiff5}).
Let us denote
\be\label{NCries9} \kappa_n = \frac{\lambda \alpha}{n},\ \ \ \Omega^I_n\ =\  \frac{\kappa_n-\sqrt{1+\kappa_n^2 }+1}{\kappa_n +\sqrt{1+\kappa_n^2  }-1}\,.\ee
Then the solution   (\ref{NCries1}) is
\be\label{NCries10} R^I_{nj}\ =\ (\Omega^I_n)^N\, F(-n,\,-N,\,2j+2;\,-2 \kappa_n (\Omega^I_n)^{-1})\,.\ee
It is integrable since $\Omega_n\,\in\,(0, 1)$ for positive $\kappa$ and under given conditions the hypergeometric function is a polynomial. The norm (\ref{Psi2}) of $R^I_{nj}$ is finite and given in terms of a generalized hypergeometric function. We do not present the corresponding cumbersome formula as it is not needed for our purposes.

\subsubsection{\it\ \textbf{ Poles of the $S$-matrix for repulsive potential}}
(These disappear from the Hilbert space of the physical states in the limit $\lambda \rightarrow 0$ )
\[ p^\lambda_n\,=\,i\,\frac{\alpha }{n}\,,\ \ \ \alpha < 0\ \ \ \Leftrightarrow\ \ \ E^{II}_{\lambda \, n}\,=\,\frac{1}{\lambda^2}\, \left(1\,+\,\sqrt{1+(\lambda\alpha/n )^2 } \right)\,>2/\lambda^{2} \]
\be\label{NCries11}  n\, =\, j+1,\ j+2,\ \dots \ee
Now (\ref{NCries1}) has the form
\be\label{NCries12} R^{II}_{nj}\ =\ (-\Omega^{II}_n)^N\, F(-n,\,-N,\,2j+2;\,2 \kappa_n (\Omega^{II}_n)^{-1} \,) \ee
where
\be
\Omega^{II}_n\ =\  -\frac{\kappa_n+\sqrt{1+\kappa_n^2 }+1}{\kappa_n -\sqrt{1+\kappa_n^2  }-1}\,.\ee
The definition of $\kappa_n$ is the same as in (\ref{NCries9}) (note that it is negative
this time). Since $\Omega^{II}_n = \Omega^I_n \in (0,1)$  the solution (\ref{NCries1})
is integrable because the hypergeometric function terminates like in the
previous case.

\section{Comments and conclusions}

The aim of this section is to sum up what has been done, compare QM and NCQM results, and finally outline what could be done in the future:
\\
We carefully defined the NC rotationally invariant analog of the
QM configuration space and the Hilbert space of operator wave
functions . The central point of our
construction was the definition of $\Delta_\lambda$,  the NC analog
of Laplacian, supplemented by a consequent definition of the
weighted Hilbert-Schmidt norm and a definition of the Coulomb
potential satisfying NC Laplace equation.
\newline
Then  we were able to  introduce the NC analog of the Schr\"{o}dinger equation, taking advantage of the spherical symmetry of the problem when separating the radial part. The knowledge of how are normal ordered powers of the "NC radial variable" correlated with the "usual" powers enabled us to solve the NC problem using the associated ordinary differential equation.
\\
Now let us see how NCQM matches the \, "standard" \, quantum mechanics.  We are sorry to bother the reader with the  QM facts which are undoubtedly familiar to them, but there seems to be no better way to compare the theories.
\\
The quantity labeling the solutions of Schr\"{o}dinger equation is energy. Some of the labels are excluded in the sense
that they cannot be attributed to a physical state. In standard QM, the solutions with negative energy were dismissed as
lacking the physical interpretation, except for those with some special energy values for which the wave function was normalizable.
One would simply expect a particle with negative energy to be trapped in certain region, and, on the contrary, since a particle
with positive energy can be pretty much anywhere, it is quite acceptable that Schr\"{o}dinger equation does not provide us
with normalizable solutions for a given case.\\
\\
The state of affairs seems to be a bit different in NCQM. Although the energy spectrum for an electron trapped in the atom
is predicted in agreement with QM with small correction of order $\lambda^2$ , there  are \textit{two} special values
of energy, $E=0$ and $E_{crit}=2\hbar^2/(m\lambda^2)$ with the following feature: certain energy values  below $E=0$  for an attractive Coulomb potential and certain values above $E_{crit}$  for repulsive potentials provide a normalizable state.
\\
There is a remarkable symmetry between normalizable states corresponding
to $E^I_n < 0$ and those corresponding to $E^{II}_n > E_{crit}$:
\\
The bound state energies are symmetric with respect to the energy
$\frac{1}{2} E_{crit}\,=\,\hbar^2/(m \lambda^2)$,
$$ E^{I,\,II}_{\lambda\,n}\ =\ \frac{1}{2} E_{crit}\left( 1\,\mp\,
\sqrt{1 + (\lambda\alpha/n)^2}\right)\,. $$
Moreover, the corresponding radial wave functions are equal up to the
change $\alpha\,\to\,-\alpha$ and the sign changing factor at each step
$\lambda$ in the radial direction,
$$ R^{II}_{nj}(-\alpha)\ =\ (-1)^N\ R^I_{nj} (\alpha)\,. $$
The same symmetry can be seen for scattering states for energies:
$$ E^I\,=\,\frac{1}{2} E_{crit}\,-\,\varepsilon\,,\ \ \ \
p^I\,=\,+\,\sqrt{\frac{1}{2} E_{crit} - \lambda^2}\,, $$
$$ E^{II}\,=\,\frac{1}{2} E_{crit}\,+\,\varepsilon\,,\ \ \ \
p^{II}\,=\,-\,\sqrt{\frac{1}{2} E_{crit} - \lambda^2}\,, $$
with $\varepsilon\,\in\,(0, \frac{1}{2} E_{crit})$. Namely,
$$ R^{II}_{\varepsilon j}(-\alpha)\ =\
(-1)^N\ R^I_{\varepsilon j} (\alpha)\,. $$
This relation follows directly from
$$ \frac{p^I + i\lambda E^I}{p^I - i\lambda E^I}\ =\
-\,\frac{p^{II} + i\lambda E^{II}}{p^{II} - i\lambda E^{II}}\ =\
\frac{\sqrt{1+\lambda^2 \varepsilon}\,+\,i\sqrt{1+\lambda^2 \varepsilon}}{
\sqrt{1+\lambda^2 \varepsilon}\,-\,i\sqrt{1+\lambda^2 \varepsilon}} $$
It would be highly desirable to see the background of those almost perfect
reflection symmetry in the energy with respect to $\frac{1}{2} E_{crit}$.
\\
\\
The reader may wonder how big the NC  corrections actually are. The answer is, that the  parameter $\lambda$ is not fixed within our
model. However, it can be estimated by some other physical
requirement. For example, one can postulate, as was done in early
days of modern physics, that the rest energy $mc^2$ of electron is
equal to the electrostatic energy of its Coulomb field. In
${\bf{R}^3_\lambda }$ this means:
\be\label{ncradius} mc^2\ =\ \frac{4\pi
\lambda^3}{8\pi}\,\mbox{Tr}\,[ (N+1)\, E^2_j ]\,\ee
where
\be E_j\ =\ \frac{e^2}{\lambda^3}\,\frac{1}{N
(N+1)(N+2)}\,x_j\,,\ee
is the NC electric field strength corresponding to NC Coulomb
potential  (the details will be published, see \cite{KP}).

We stress that in the NC case the electrostatic energy of
electron, determined by the trace in (\ref{ncradius}), is finite
(no cut-off at short distance is needed). A straightforward
calculation of the trace in (\ref{ncradius}) gives the relation:
\be mc^2\ =\ \frac{3}{8}\,\frac{e^2}{\lambda}\ \ \ \Rightarrow\ \
\ \lambda\ =\ \frac{3}{8}\,\frac{e^2}{mc^2}\,\equiv\,\lambda_0
\,.\ee
This $\lambda_0$ is fraction of the classical radius of electron
$r_0 = e^2/mc^2$: $\lambda_0 = 1.06\,\times\,10^{-15}\,m = 1.06\,
fm$ (the coincidence with the proton radius is purely accidental).

The NC corrections to the $H$-atom energy levels given in this paper
are of  order $(\lambda_0/a_0)\, =\, (9/64)\,
\alpha^2_0\,\approx\, 4\,\times\,10^{-11}$ (here $\alpha_0 \approx
1/137$ is fine structure constant). Such tiny corrections to
energy levels are beyond any experimental evidence. Moreover, at
$\lambda_0 \approx 1 \,fm$ relativistic and QFT effects become
essential.

Our investigation indicates that the noncommutativity of the
configuration space is fully consistent with the general QM
axioms, at least for the  $H$-atom bound states. However, a more
detailed analysis of the Coulomb problem in ${\bf{R}^3_\lambda}$
would be a desirable dealing, e.g, with the following aspects:

$\bullet$ dyon problem (electron in
the electric point charge and magnetic monopole field), Pauli
$H$-atom (non-relativistic spin);

$\bullet$ Coulomb problem in ${\bf{R}^3_\lambda}$ and its dynamical
symmetry, QM supersymmetry and integrability of the Coulomb
system.
\\
Besides non-relativistic $H$-atom, there are other systems that
would be interesting to investigate within NC configuration space
${\bf{R}^3_\lambda}$, e.g. Dirac $H$-atom (relativistic
invariance?), non-Abelian monopoles, or spherical black-holes.

\newpage %\vskip1cm

{\bf\Large Appendix A}\vskip0.5cm

Here we prove two formulas (\ref{appen}) we need for the
calculation of the NC Coulomb Hamiltonian. (We skip indices $j$ and
$m$ here.)

a. Let us begin with the first formula:
\beqa\nn [a^\dagger_\alpha ,[a_\alpha
,\Psi_{jm}]] &=& \lambda^j [a^\dagger_\alpha
,\,[a_\alpha ,\sum_{(jm)}\frac{(a^\dagger_1
)^{m_1}\,(a^\dagger_2)^{m_2}}{m_1!\,m_2!}\ :{\cal R}:
\frac{a^{n_1}_1\,(-a_2)^{n_2}}{n_1!\ n_2!}]]\\
\nn &=&\ \lambda^j \sum_{(jm)}[a_\alpha
,\frac{(a^\dagger_1 )^{m_1}
(a^\dagger_2)^{m_2}}{m_1!\,m_2!}]\,
[a^\dagger_\alpha,:{\cal R}:]\,
\frac{a^{n_1}_1(-a_2)^{n_2}}{n_1!\ n_2!} \\
\nn &+&\ \lambda^j \sum_{(jm)}[a_\alpha
,\,\frac{(a^\dagger_1
)^{m_1}\,(a^\dagger_2)^{m_2}}{m_1!\,m_2!}]\,:{\cal R}:\,
[a^\dagger_\alpha,\,\frac{a^{n_1}_1
(-a_2)^{n_2}}{n_1!\ n_2!}] \\
\nn  &+& \lambda^j \sum_{(jm)}\frac{(a^\dagger_1
)^{m_1}\,(a^\dagger_2)^{m_2}}{m_1!\,m_2!}\,
[a^\dagger_\alpha,\,[a_\alpha ,:{\cal R}:]]\,
\frac{a^{n_1}_1\,(-a_2)^{n_2}}{n_1!\
n_2!} \\
%\nn
\label{A1} &+& \lambda^j \sum_{(jm)}\frac{(a^\dagger_1
)^{m_1}\,(a^\dagger_2)^{m_2}}{m_1!\ m_2!}\ [a_\alpha
,:{\cal R}:]\, [a^\dagger_\alpha,\frac{a^{n_1}_1
\,(-a_2)^{n_2}}{n_1!\ n_2!}] ,  \eeqa
where ${\cal R}\,=\,\sum_{k=0}^\infty\,c_k \varrho^k =\,\sum_{k=0}^\infty\,c_k \lambda^k N^k$ .
Now we shall use the
following commutation relations
%\nn
\beqa\nn [a^\dagger_\alpha,:N^k:]\,=- \,k\,
a^\dagger_\alpha\, :N^{k-1}:\ \ &\Rightarrow& \ \
[a^\dagger_\alpha,:{\cal R}:]\,=\, -\lambda a^\dagger_\alpha\,
:\partial_{\varrho}{\cal R}:\,,\\
%\nn
\label{A2 } [a_\alpha,:N^k:]\,=\, k\,
:N^{k-1}:\,a_\alpha\ \ &\Rightarrow& \ \
[a_\alpha,:{\cal R}:]\,=\,\lambda :\partial_{\varrho}{\cal R}:
\,a_\alpha \,,\eeqa
where $\partial_{\varrho}$ denotes the derivatives with respect to
$\varrho$: $\partial_{\varrho}{\cal R}\,=\, \sum_{k=1}^\infty
k\,c_k\, \varrho^{k-1}$.

It is easy to see that the second line in (\ref{A1}) vanish, and
the the first and third line give the same contribution
\be\label{A3 }  \sum_{(jm)}\,\frac{(a^\dagger_1
)^{m_1}\,(a^\dagger_2)^{m_2}}{m_1!\,m_2!}\
(-\lambda \, j\,:\partial_{\varrho}{\cal R}:)\
\frac{a^{n_1}_1\,(-a_2)^{n_2}}{n_1!\ n_2!}\,.\ee
From (\ref{A2 }) the double commutator
$[a^\dagger_\alpha,[a_\alpha,:{\cal R}:]]$ follows
directly, and this gives the value of the third line in (\ref{A1})
\be\label{A4 } \sum_{(jm)}\,\frac{(a^\dagger_1 )^{m_1}\,
(a^\dagger_2)^{m_2}}{m_1!\,m_2!}\ (-\lambda :\varrho \,
\partial^2_{\varrho} {\cal R}:\,+\,2 \lambda \,:\partial_{\varrho}{\cal R}:)\
\frac{a^{n_1}_1\,(-a_2)^{n_2}}{n_1!\ n_2!}\,.\ee
The last two equations yields the first
formula in (\ref{appen}).

b. From equation (\ref{nk}) it follows easily
\be\label{A5} N\,:N^k:\ =\ :N^{k+1}:\,+\,k\,:
N^k:\ \ \ \Rightarrow\ \ \ \varrho\,:{\cal R}:\ =\
:\varrho \,{\cal R}:\,+\,\lambda :\varrho\,\partial_{\varrho}{\cal R}:\,.\ee
This relation gives directly the second formula in (\ref{appen}) :
\[ (N +1)\,\sum_{(jm)}\,\frac{(a^\dagger_1
)^{m_1}\,(a^\dagger_2)^{m_2}}{m_1!\,m_2!}\ :{\cal R}:\
\frac{a^{n_1}_1\,(-a_2)^{n_2}}{n_1!\ n_2!}\]
\beqa\nn &=& \ \sum_{(jm)}\,\dots\ [(N +j +1)\,:{\cal R}:]\
\dots \\
\label{A6} &=& \ \sum_{(jm)}\,\dots\ :[(N +j
+1)\,{\cal R}\,+\,\varrho\,\partial_{\varrho}{\cal R}]:\
\dots\,,\eeqa
where we have replaced  both untouched factors containing
annihilation and creation operators by dots.

\vskip1cm {\bf\Large Appendix B} \vskip0.5cm

This part of the Appendix deals with how are normal and usual powers of $\varrho$ correlated, and how can this information be used to rewrite various kinds of power series. So the key  relation of this section relates $:\varrho^n:$ and $\varrho^n$.
\be
\begin{array}{lr}
:\varrho^k: \ = \lambda^k \frac{N!}{(N-k)!}=(-\lambda)^k(-N)_k \ \ \  & \ \ \  :\varrho^{-k}: \ = \lambda^{-k} \frac{N!}{(N+k)!}= \lambda^{-k}
\frac{1}{(N)_k}\\
\end{array}
\ee
$(N)_k$ stands for Pochhammer symbol. Since $\varrho = \lambda N$, there really are  powers of  $\varrho$ on the right-hand sides. \\
Now to the above mentioned power series. Let us start with finding out how does the function $:e^{\beta \varrho}:$ modify when we rewrite  $:\varrho^n:$  in terms of  $\varrho^n$  in the corresponding Taylor series  ($\beta$
denotes some arbitrary constant here):
\be
\begin{array}{cl}
:e^{\beta \varrho}: & = \sum^\infty_{k=0} \frac{\beta^k}{k!}:\varrho^k:  \ = \sum^\infty_{k=0} \frac{(\beta\lambda)^k}{k!}
\frac{N!}{(N-k)!} = \\
& \\
& = \left(1+ \lambda \beta \right)^{N} = \left(  1+ \lambda \beta \right)^{\frac{\varrho}{\lambda}} \\
\end{array}
\ee
Considering the limit $\lambda\rightarrow 0$  the above equation corresponds to the knowm Euler's formula. A potential doubt arising from the colon
marks on the left hand side ought to be dismissed due to the fact that $:\varrho^n: \, \rightarrow \varrho^n$ if $\lambda\rightarrow 0$.
There is no way to distinguish between the normal and usual ordering in $\lambda=0$ world. \\
\\
We can move to the more complex tasks now. In the course of many calculations we need to handle expressions of the kind $:\varrho^n e^{\beta \varrho}:$
\be\nn
\begin{array}{cl}
 :\varrho^n e^{\beta \varrho}: &= \sum^\infty_{k=0}\frac {\beta^k :\varrho ^{n+k}:}{k!}=
\sum^\infty_{k=0}\frac {\beta^k \lambda^{n+k}}{k!} \frac{N!}{(N-(n+k))!} = \\
  & \\
&=  \lambda^n  \frac{N!}{(N-n)!}    \sum^\infty_{k=0}\frac {\beta^k \lambda^{k}}{k!}
\frac{(N-n)!}{(N-n-k)!}       =    \\
& \\
  &      = \ \lambda^n  \frac{N!}{(N-n)!}   \left(   1+ \beta \lambda \right)^{N-n}
     \\
\end{array}
\ee
And in the case of negative powers:
\be\nn \begin{array}{cl}
 :\varrho^{-n} e^{\beta \varrho}: &= \sum^\infty_{k=0}\frac {\beta^k :\varrho ^{k-n}:}{k!}=
\sum^\infty_{k=0}\frac {\beta^k \lambda^{k-n}}{k!} \frac{N!}{(N-(k-n))!} = \\
  & \\
&=  \lambda^{-n}  \frac{N!}{(N+n)!}    \sum^\infty_{k=0}\frac {\beta^k \lambda^{k}}{k!}
\frac{(N+n)!}{(N+n-k)!}       =    \\
& \\
  &      = \  \lambda^{-n}  \frac{N!}{(N+n)!}    \left(   1+ \beta \lambda \right)^{N+n}    \\
\end{array}
\ee
That is almost all that is needed to get rid of the normal powers in the solutions of the NCQM equations. Since hypergeometric functions are
often written in the form of power series with coefficients expressed in terms of the Pochhammer symbols, it is useful to take notice of  the following identities when it comes to handling factorials:
\be\nn \begin{array}{ll}
(a)_n &= a (a+1)...(a+n-1)=\frac{(a+n-1)!}{(a-1)!}\\
  &\\
  \frac{N!}{(N+n)!} &= \frac{N!}{(N+n)(N+n-1)...(N+1)N!}= \frac{1}{(N+1)_n }\\
  & \\
\frac{N!}{(N-n)!} &= \frac{N(N-1)...(N-n+1)(N-n)!}{(N-n)!}= (-1)^n (-N)(-N+1)...(-N+n-1)=\\
  & =(-1)^n (-N)_n\\
\end{array}
\ee
Now we are going to supply the reader with some of the  calculations which were skipped in the previous sections. To keep reasonable length of the formulas, we will replace the long expressions with $a,\,c, \, \beta, \, Q$...substituting the "right arguments" instead of our abbreviations and completing the calculation that way is straightforward...yet not so tempting.
\\
The following  derivation was  left out in (\ref{reg3}):
\be
\begin{array}{lcl}
 &:e^{\beta \varrho} \phi\left( a; c; Q\varrho\right):&= \, :e^{\beta \varrho}
 \sum^\infty_{m=0}\frac{(a)_m}{(c)_m}\frac{(Q\varrho)^m}{m!}:   \\
           & = & \,  \sum^\infty_{m=0}\left[ 1+\lambda\beta\right]^{N-m}\frac{(a)_m}{(c)_m}(Q\lambda)^m\frac{N!}{(N-m)!m!}   \\
          &= & \left[ 1+\lambda\beta\right]^{N}  \sum^\infty_{m=0}\left[ \frac{Q\lambda}{
          1+\lambda\beta}\right]^{m}\frac{(a)_m}{(c)_m}\frac{1}{m!}(-1)^{m}(-N)_m  \\
                 & = & \left[ 1+\lambda\beta\right]^{N}  \sum^\infty_{m=0}\left[ \frac{-Q\lambda}{ 1+\lambda\beta}\right]^{m} \frac{(a)_m(-N)_m}{(c)_m
         }\frac{1}{m!}  \\
                   &= & \left[ 1+\lambda\beta\right]^{N}  F\left(a,-N; c;  \frac{-Q\lambda}{ 1+\lambda\beta}\right)
\end{array}
\ee
This is to be done and suitably used if one wishes to rewrite (\ref{reg3}) in terms of the fundamental system (\ref{y5-7}) and to get rid of the normal ordering only thereafter:
\be\label{psiodb}
\begin{array}{lcl}
&:e^{\beta \varrho} \psi\left( a; c; Q\varrho \right):&= \, :e^{\beta\varrho} \times \\
                           & & \times \sum^\infty_{m=0}\frac{(-1)^{m}(a)_m(a-c+1)_m (Q\varrho)^{-a-m}}{m!}:   \\
           & = & \left[ 1+\lambda\beta\right]^{N}\left[\frac{1+\lambda\beta}{\lambda Q}\right]^a \frac{N!}{(N+a)!} \times \\
              && \times \sum^\infty_{m=0} \frac{(a)_m(a-c+1)_m }{m!}  \left[\frac{1+\lambda\beta}{-\lambda Q}\right]^m \frac{(N+a)!}{(N+m+a)!} \\
                    & =& \left[ 1+\lambda\beta\right]^{N}\left[\frac{1+\lambda\beta}{\lambda Q}\right]^a \frac{N!}{(N+a)!} \times \\
                    && \times \sum^\infty_{m=0} \frac{(a)_m(a-c+1)_m }{(N+a+1)_m m!}  \left[\frac{1+\lambda\beta}{-\lambda Q}\right]^m  \\
           & = & \left[ 1+\lambda\beta\right]^{N}\left[\frac{1+\lambda\beta}{\lambda Q}\right]^a \frac{N!}{(N+a)!}\times \\
            && \times F\left( a,\, a-c+1,\, N+a+1, \,
          \frac{1+\lambda\beta}{-\lambda Q}\right)  \\
\end{array}
\ee
To complete (\ref{gsol2}), this was needed:
\be
\begin{array}{lcl}
 &:\varrho^{-j-1/2} J_{-2j-1}(\sqrt{8\alpha \varrho}):&= \, \sum^\infty_{m=0}\frac{(-1)^{m+1}
 (2\alpha)^{m+j+1/2}}{(2j+1+m)!m!}:\varrho^{m}:   \\
 &=&- \frac{(2\alpha)^{j+1/2}}{(2j+1)!}\sum^\infty_{m=0}\frac{(-1)^{m+1}  (2\alpha\lambda)^{m}(2j+1)!N!}{(2j+1+m)!m!(N-m)!}    \\
   &=&- \frac{(2\alpha)^{j+1/2}}{(2j+1)!}\sum^\infty_{m=0}\frac{ (-N)_m (2\alpha\lambda)^{m}}{(2j+2)_m m!}    \\
      &=&- \frac{(2\alpha)^{j+1/2}}{(2j+1)!} \phi(-N, \, 2j+2, \,  2\alpha\lambda)    \\
\end{array}
\ee
The equation (\ref{gsol3}) required:
\be
\begin{array}{lcl}
 &:e^{-2\varrho/\lambda}\varrho^{-j-1/2} J_{-2j-1}(\sqrt{8\alpha\varrho}):&= \, \sum^\infty_{m=0}\frac{(-1)^{m+1}
 (2\alpha)^{m+j+1/2}}{(2j+1+m)!m!}:e^{-2\varrho/\lambda}\varrho^{m}:   \\
 &=&- \frac{(2\alpha)^{j+1/2}}{(2j+1)!}\sum^\infty_{m=0}\frac{(-1)^{m}(1-2)^N  (2\alpha\lambda)^{m}(2j+1)!N!}{(1-2)^m (2j+1+m)!m!(N-m)!}
          \\
           &=&\frac{(2\alpha)^{j+1/2}(-1)^{N+1}}{(2j+1)!}\sum^\infty_{m=0}\frac{ (-N)_m (-2\alpha\lambda)^{m}}{(2j+2)_m m!} \\
            &=&\frac{(-1)^{N+1}(2\alpha)^{j+1/2}}{(2j+1)!} \phi(-N, \, 2j+2, \,  -2\alpha\lambda)    \\
\end{array}
\ee

\vskip1cm {\bf\Large Appendix C} \vskip0.5cm
In this section we are going to look on the scattering case more closely, namely the part dealing with decomposition of the solution (\ref{NCries1}) into two parts corresponding to the in- and out- going spherical wave. Firstly let us remind the form in which we got  (\ref{NCries1}):
\be\nn R_{E j} =\left( \frac{p+i\lambda E}{p-i\lambda E}\right)^N\,F\left(j+1-i\frac{\alpha }{p},\,-N,\,2j+2;\, 2i\lambda p\, \frac{p-i\lambda E}{p+i\lambda E}\right)\,\ee
If we , for the sake of brevity, denote
\be\label{subst}
\begin{array}{llll}
a=j+1-i\frac{\alpha}{p} &, \,\,\, b=-N &, \,\,\, c=2j+2 &, \, \,\, z=2i\lambda p \frac{p-i\lambda E}{p+i \lambda E} , \\
\end{array}
\ee
then $R_{Ej}$ can be written as
\be R_{E j} = \left( 1-z\right)^{b/2}\,F\left(a,\,b,\,c;\, z\right)\,\ee
According to Kummer identities (see \cite{Bat}), $F(a,b,c; z)$ can be written as a linear combination of two other solutions of the hypergeometric equation (\ref{hypgeoeq}), namely
\beqa\label{FF}
\nn &(-z)^{-a}F(a,\, a+1-c, \, a+1-b; \,z^{-1} ) & \\
\nn &\mbox{\,\,and \,\,}& \\
 &(z)^{a-c}(1-z)^{c-a-b}F(c-a,\, 1-a, \, c+1-a-b; \,z^{-1}(z-1) ) &
\eeqa
Take notice of the fact that if we  decompose our radial solution (\ref{reg3}) before handling the normal ordering, i.e. if we write the confluent hypergeometric function (\ref{y1}) in terms of the fundamental system (\ref{y5-7}) of confluent hypergeometric equation (\ref{confl}), and deal with the normal ordering  only thereafter (see Appendix B, eq. (\ref{psiodb})), we will  end up with (\ref{FF}) again.
So after using one of the numerous Kummer relations listed in \cite{Bat}, we have
\be
\begin{array}{lll}
R_{Ej} & =& \left( 1-z\right)^{b/2}\,F\left(a,\,b,\,c;\, z\right) = \\
        & & \\
   & =&  e^{i\pi(c-a)}\frac{\Gamma(c)\Gamma(1-b)}{\Gamma(a)\Gamma(c+1-a-b)} \times\\
   & & \times   (z)^{a-c}(1-z)^{c-a-b/2}F(c-a,\, 1-a, \, c+1-a-b; \,z^{-1}(z-1) ) +\\
   $$ \\
   & +&   e^{i\pi(1-a)}     \frac{\Gamma(c)\Gamma(1-b)}{\Gamma(c-a)\Gamma(a+1-b)}   \times \\
   & & \times (-z)^{-a}\left( 1-z\right)^{b/2} F(a,\, a+1-c, \, a+1-b; \,z^{-1} ) \\
\end{array}
\ee
Substituting back (\ref{subst}) leads to
\be\label{rozklad1}
\begin{array}{lll}
R_{Ej} & =& \left( \frac{p+i\lambda E}{p-i\lambda E}\right)^N\,F\left(j+1-i\frac{\alpha }{p},\,-N,\,2j+2;\, 2i\lambda p\, \frac{p-i\lambda E}{p+i\lambda E}\right)\\
        & & \\
   & =&  e^{i\pi(j+1+i\frac{\alpha}{p})}\frac{\Gamma(2j+2)\Gamma(N+1)}{\Gamma(j+1-i\frac{\alpha}{p})\Gamma(j+2+i\frac{\alpha}{p} +N)}
   \left[2i\lambda p\, \frac{p+i\lambda E}{p-i\lambda E}\right]^{-j-1-i\frac{\alpha}{p}} \left(\frac{p-i\lambda E}{p+i\lambda E}\right)^N\times\\
   & & \times F\left(j+1+i\frac{\alpha}{p} ,\,\,\,-j+i\frac{\alpha}{p},\,\,\,\frac{r}{\lambda} +j+2+i\frac{\alpha}{p};\,\,\,-\frac{1}{2i\lambda p}\, \frac{p-i\lambda E}{p+i\lambda E}\right) \\
    & &  \\
      & +& e^{i\pi(-j-1+i\frac{\alpha}{p})}\frac{\Gamma(2j+2)\Gamma(N+1)}{\Gamma(j+1+i\frac{\alpha}{p})\Gamma(j+2-i\frac{\alpha}{p} +N)}
   \left[-2i\lambda p\, \frac{p-i\lambda E}{p+i\lambda E}\right]^{-j-1+i\frac{\alpha}{p}} \left(\frac{p+i\lambda E}{p-i\lambda E}\right)^N\times\\
   & & \times F\left(j+1-i\frac{\alpha}{p} ,\,\,\,-j-i\frac{\alpha}{p},\,\,\,\frac{r}{\lambda} +j+2-i\frac{\alpha}{p};\,\,\,\frac{1}{2i\lambda p}\, \frac{p+i\lambda E}{p-i\lambda E}\right) \\
   \end{array}
\ee
Keep in mind that our endeavour is to compare the NCQM result with the QM one. It is therefore suitable  to rewrite the prefactors in front of the hypergeometric functions as exponentials. This way we obtain, besides other factors, also logarithms of $\Gamma$-functions, which can be rewritten using the following formula (see \cite{Bat}):
\be\label{lngamma}
\ln \left(\Gamma (z+a) \right)=\left(z+a-\frac{1}{2}\right)\ln(z) - z + \frac{1}{2} \ln(2\pi)+ \sum^{\infty}_{n=1} \frac{B_{n+1}(a)}{n(n+1)}z^{-n} \ee
Here $B_{n}(a)$ is a Bernoulli polynomial. After certain rearrangements we finally acquire the result (\ref{rozklad}).

\end{document}